\documentclass[useAMS,usenatbib]{mn2e}
\usepackage{graphics}
 \usepackage{graphicx}
 \usepackage{float}
\usepackage{placeins}
\usepackage{color}

\title[Survival of impactors in hypervelocity collisions]{Survival of the impactor during hypervelocity collisions I:\\ An analogue for low porosity targets.}
\author[C. Avdellidou et al.]{C. Avdellidou$^{1}$\thanks{E-mail:
ca332@kent.ac.uk}, M.C. Price$^{1}$, M. Delbo$^{2}$, P. Ioannidis$^{3}$ and M.J.Cole$^{1}$\\
$^{1}$Centre for Astrophysics and Planetary Science, School of Physical Sciences, University of Kent, Canterbury, CT2 7NH, UK\\
$^{2}$Laboratoire Lagrange, Universit\'e C\^ote d'Azur, Observatoire de la C\^ote d'Azur, CNRS,
Blvd de l'Observatoire, CS 34229,\\ 06304 Nice cedex 4, France\\
$^{3}$Hamburger Sternwarte, Universit\"{a}t Hamburg, Gojenbergsweg 112, 21029, Hamburg, Germany\\}
\begin{document}

\date{Accepted . Received }

\pagerange{\pageref{firstpage}--\pageref{lastpage}} \pubyear{2002}

\maketitle

\label{firstpage}

\begin{abstract}
Recent observations of asteroidal surfaces indicate the presence of materials that do not match the bulk lithology of the body. A possible explanation for the presence of these exogenous materials is that they are products of inter-asteroid impacts in the Main Belt, and thus interest has increased in understanding the fate of the projectile during hypervelocity impacts. In order to gain insight into the fate of impactor we have carried out a laboratory programme, covering the velocity range of 0.38 -- 3.50 km/s, devoted to measuring the survivability, fragmentation and final state of the impactor. Forsterite olivine and synthetic basalt projectiles were fired onto low porosity ($\textless$10\%) pure water-ice targets using the University of Kent's Light Gas Gun (LGG). We developed a novel method to identify impactor fragments which were found in ejecta and implanted into the target. We applied astronomical photometry techniques, using the SOURCE EXTRACTOR software, to automatically measure the dimensions of thousands of fragments. This procedure enabled us to estimate the implanted mass on the target body, which was found to be a few percent of the initial mass of the impactor. We calculated an order of magnitude difference in the energy density of catastrophic disruption, Q*, between peridot and basalt projectiles. However, we found very similar behaviour of the size frequency distributions for the hypervelocity shots ($\textgreater$1 km/s). After each shot, we examined the largest peridot fragments with Raman spectroscopy and no melt or alteration in the final state of the projectile was observed.

\end{abstract}


\section{Introduction}
\label{intro}
Impacts have shaped the asteroids, and their size frequency distribution, over 4.5 billion years of Solar System evolution \citep{2005Icar..179...63B} and are responsible for the formation of asteroid families. The appearance and morphology of asteroidal surfaces are also the result of impact processes, which are responsible for the formation of craters and the production of regolith \citep{1997M&PS...32..179H} (although it has been recently shown that the regolith can be efficiently produced by thermal fragmentation of surface rocks by \cite{2014Delbo}).
Over the last four decades, a plethora of laboratory experiments and computer simulations have provided insights into collisional processes that constitute the foundation of our current understanding of large-scale asteroid collisions \citep{2002holsappleast3}. The majority of these studies focused on the fate of the target after an impact (e.g. degree of fragmentation, catastrophic disruption of different materials, crater sizes etc). They have provided data on the speed and size distributions of the fragments using several target materials, mostly cement mortar, basalt or ice, while the projectiles are mostly iron, copper, pyrex or basalt. 
Furthermore, efforts have been devoted to the study of the mass and the velocities of the ejecta \citep{2011Housen, 2007michikami}.

However, the fate of the impactor at impact speeds of a few km/s is still poorly understood. The investigation of the projectile, and projectile debris, during hypervelocity impacts is crucial to explain the observations of mixed mineralogies on the surface of asteroids. Such phenomena, which have been observed only relatively recently, are the source of the olivine and dark material deposits observed on Vesta \citep{2012Natur.491...83M, 2012Icar..221..544R} and probably of the "Black Boulder" on (25143) Itokawa \citep{2011LPI....42.1821H}. Mixing of asteroid material with different lithology through impacts is also necessary to explain the nature of the Near-Earth asteroid 2008 TC$_{3}$, a multi-lithology body whose formation mechanism is still not completely understood \citep{2009Natur.458..485J, 2010MAPS...45.1638B}. 2008 TC$_{3}$ impacted Earth's atmosphere  on October 7, 2008 and it is estimated that it exploded approximately 37 km above the Nubian Desert in Sudan.
A large number ($\sim$600) of small (0.2 -- 379 g) meteorites were recovered from 2008 TC$_{3}$, and are collectively called Almahata Sitta. The big surprise was that those meteorites were of various mineralogical types: analysis of 110 meteorites revealed 75 ureilites, 28 enstatite chondrites (both EH and EL), 5 ordinary chondrites (H, L, LL), one carbonaceous chondrite (CB) and one which is a previously unknown type of chondrite related to R-chondrites. This fact has changed completely our paradigm that one meteorite fall produces meteorites of only one particular type.
A recent study by \cite{2012MNRAS.424..508G} has shown that there is a small probability that foreign material remains on the surface of a body after low speed collisions. 
However, their results were based on the assumption that, in order to preserve the impactor, an impact velocity $\leq$ 0.5 km/s is required, which is much smaller than the typical impact velocity (about 5 km/s) among random asteroids \citep{1994Icar..107..255B, 2011obrien}.
The second confirmed case of heterogeneous meteorite is Benesov \citep{2014benesov}. Surprisingly, one meteorite was H chondrite, one was LL chondrite and one was LL chondrite with embedded achondritic clast. 
These findings shed new light on some old meteorite finds, such as the Galim meteorite fall (LL+EH), Hajmah (ureilite+L), Gao-Guenie (H+CR), and Markovka (H+L) \citep{2015borovicka}. 
Therefore asteroids with mixed mineralogies might be more abundant than previously thought, but their formation mechanism(s) remain mysterious \citep{2014horstmann}. 
One possible solution is that the heterogeneous composition of some asteroids was inherited from a time when the asteroid belt was in a different dynamical state, most likely in the very early Solar System.

These findings call for new experiments devoted to ascertaining what is the highest velocity that projectile material can be preserved and/or implanted onto asteroids via impacts. Pioneering experiments  by \cite{1984schultz} and \cite{1990schultzgsa} demonstrated a change in projectile fragmentation and cratering efficiency as a function of impact velocity. Recently,
\cite{2014M&PS...49...69N} performed several laboratory experiments using pyrophyllite and basalt projectiles fired onto regolith-like sand and aluminum targets. They found that projectile material survived the impacts, although the degree of fragmentation of the projectile depended on the impact energy (Q) and the strength of the projectile, along with the strength and the porosity of the target.
However, considering an average impact speed of v = 5.3 km/s for Main Belt asteroid collisions \citep{1994Icar..107..255B}, the collisional speed range that was tested (\textless 1 km/s) in the experiments of \cite{2014M&PS...49...69N} was at the lower end of inter-asteroid collision velocities. 
Moreover, \cite{Daly2013, Daly2014, Daly2015}, \cite{2016Daly} and, \cite{Dalynew} used aluminum and basalt projectiles  which were fired onto pumice and highly porous water-ice trying to explain the implantation of an impactor's material onto vestan regolith, and the possibility of a similar process onto Ceres' surface. McDermott et al. (in preparation) used copper projectiles impacting porous ($\sim$50\%) water-ice targets at a wide range of speeds (1.00 -- 7.05 km/s). Their results show that the projectile can be recovered completely intact at speeds up to 1.50 km/s, whereas it started to break into smaller fragments at speeds above 1.50 km/s. Increasing the impact velocity was found to produce an increasing number of projectile fragments of decreasing size.
All of these recent investigations try to shed light onto the fate of the impactor. The main question that is addressed is how much of the impactor's material is embedded on/into the target by using different combinations of materials, trying to simulate collisions in the Main Belt and on the surface of icy bodies.
While the experiments of \cite{2014M&PS...49...69N} were limited to speeds \textless 1km/s, the McDermott et al. study used only porous water-ice and a copper projectile - which is an atypical type of impactor material in the Solar System.

In this work we advance on the investigation of the fate of the projectile during hypervelocity impacts by firing lithological projectiles, olivine and basalt, onto low-porosity water-ice targets, at a wide range of speeds between 0.38 -- 3.50 km/s, and by using novel methods to detect and measure sizes of impactor fragments down to a size-scale of a few microns. 

The structure of the paper is the following:
In Section \ref{experiments} we give a description of the materials that were used and the set-up of the experiments, along with a description of the method we established to detect, measure and analyse the impactors' fragments found both as ejecta and implanted onto/into the target. We derive the projectile fragments' volume, mass and size distribution. In Section \ref{results} we present the results of the experiments and the calculation  from hydrocode modelling of the pressures and temperatures at the time of the impact and, additionally, describe the state of the recovered fragments of the projectile. Finally we discuss our results and give the implications for impacts to induce lithological mixing on asteroids.

\section[]{Methods}
\label{experiments}

%
Our methodology consists of six steps: 
\begin{enumerate}
\item{We start by carrying out a physical characterisation of the projectiles pre-shot; namely we measure their sizes, masses and perform Raman spectroscopy. Raman spectroscopy is used to identify possible Raman line shifts due to deformation of the projectile's crystal structure induced by the impact shock. }

\item{The projectiles are fired onto the ice targets. The formation of impact craters are observed, although due to the ephemeral nature of the target they were not measured. The projectile and target material, along with contaminating residues from the gun, collected by our set up (ejecta collector -- see Fig.\ref{setup}).}

\item{We collect all the projectile fragments and visually identify the largest of them. The ratio between the mass of the largest fragment to the initial mass of the projectile, gives information about the degree of fragmentation of the latter.}

\item{The ice from the target and the ejecta collector is melted and the water, plus projectile fragments, plus contaminating gun debris, is filtered.}

 \item{The filters are then analysed using a Scanning Electron Microscope (SEM).}
\item{The final phase consists in analysing the data from the SEM, discriminating projectile fragments from gun detritus and allowing us to build the size frequency distributions (SFDs) of the fragments and quantify the amount of projectile embedded in the target and its level of fragmentation.}
\end{enumerate}

In the remainder of this Section, we present our projectile and target material choice and we detail each experimental and data analysis method.  

\subsection{The projectiles}

In order to unambiguously separate projectile fragments from those of the targets and gun contamination, we used a high purity, Mg-rich olivine - in the form of a gem quality peridot - and synthetic basalt spheres as projectiles and high-purity water-ice as the target. These materials were also chosen as: 
(a) olivine is one of  the most common minerals in the Solar System. Olivine and pyroxene minerals are the primary minerals in stony and stony-iron meteorites, 75\% of chondrite meteorites and 50\% of pallasites \citep{propertiesmeteorites,2002aste.conf..183G}; (b) Mg-rich olivine (fosterite) has  been detected in spectra of several cometary tails and is present in the majority of comet Wild 2 samples returned by  NASA's {\it Stardust Mission} \citep{2006Sci...314.1735Z}; (c) parallel studies of the spectral features of the dust particles, observed in  exo-planetary system $\beta$ Pictoris \citep{2012Natur.490...74D}, confirm similar abundance of Mg-rich olivine in respective areas (large heliocentric distances) to our Solar System; (d) Fe-rich olivine (fayalite) is mostly encountered  in asteroid mineralogies \citep{2011Sci...333.1113N} and therefore in the warmer, inner parts, of  planetary space \citep{ 2012A&A...542A..90O}. Possible explanations for this distribution of the different types of olivine are: (1) the presence of water on comets which leads to aqueous alteration, as the fayalite may not survive in the presence of water \citep{ 2012A&A...542A..90O} and, (2) the higher abundance of heavier elements, such as Fe, in the inner Solar System;
(e) Basalt is considered to be the main material on the surface of the differentiated asteroids. Differentiation, which leads to a multi-layered body with core, mantle and crust and the production of basalt, occurred in the early Solar System. It is found in the basaltic eucrites and diogenites of the HED meteorites \citep{2011SSRv..163..141M} which are linked with asteroid Vesta \citep{2012Sci...336..684R};
%
%
(f) although initially it was commonly thought that basalt is associated only with the Vestoids (asteroids which share spectroscopic data and are dynamically connected with Vesta) observations have shown that V-type asteroids do also exist  in other locations in the Main Belt \citep{Moskovitz200877}. 

The peridots, roughly 3 mm in diameter, are high quality gemstone olivine in brilliant cut, with no visible inclusions or cracks. Additionally their composition was very uniform (measured using Raman and verified by quantitative Energy-dispersive X-ray spectroscopy) and found to be Mg$_{82}$Fe$_{18}$SiO$_{4}$ using the simplified equations of \cite{Foster20131} with a compositional variance across the surface of 1\%. 

The basalt projectiles, 2.0 -- 2.4 mm spheres in diameter, were not of a natural basalt rock but synthetic spheres, sourced from `Whitehouse Scientific' with a composition of SiO$_{2}$ (43\%), Al$_{2}$O$_{3}$ (14\%), CaO (13\%), Fe$_{2}$O$_{3}$ (14\%), MgO (8.5\%), Na$_{2}$O/K$_{2}$O (3.5\%) and Others (4\%).
These projectiles are homogeneous and compositionally identical and thus we maximise the  reproducibility of the shots.

The compressive strength for forsterite and basalt projectiles was taken as 80 MPa and 100 MPa respectively \citep{propertiesmeteorites}.

\subsection{The target}

For the purposes of our work, simulating collisions at laboratory scales, it is essential to know the mechanical properties (strengths and micro/macro-porosity) of small bodies. Several literature sources give the compressive and tensile strength for a series of meteorites, however the number that has been studied is very limited. \cite{2011popova} summarises data from several meteorites, giving the ranges of compressive and tensile strengths to be 20 -- 450 MPa and 2 -- 62 MPa for L ordinary chondrites, and 77 -- 327 MPa and 26 -- 42 MPa for H ordinary chondrites. However, the calculated bulk strengths during entry of similar type meteoroids into the Earth's atmosphere are much lower than the above-quoted strengths. 
The average meteorite microporosity for the different types of ordinary and carbonaceous chondrites ranges between 6 -- 16\%. However, only the very largest asteroids seem to have comparable bulk porosity with their equivalent meteorite microporosity. The average bulk porosity for the S-type asteroids is $\sim$30\%, while for C-types is around $\sim$40\% \citep{2002BRITTast3}.
As the porosity of a body increases, the strength decreases, which could be an explanation of the big difference between the calculated and observed strength of bolides. This may explain the high altitude where some meteoroids start to disrupt and also the greater abundance of ordinary chondrites compared to carbonaceous chondrites among the meteorite samples.

In order to start our study and investigate a range of porosities and strengths we used a high purity water-ice target of low porosity, comparable to the microporosities of the examined meteorites. This was also chosen because one of the main aims of this study was to attempt to recover projectile fragments within the target. By using a water-ice target, the target only had to melt and the resulting water filtered through clean, 0.1 $\umu$m pore-size filters to recover projectile fragments.
Each water-ice target was prepared and frozen (following an identical procedure for each shot) down to $-130${$^{\circ}$}C, before being placed into the target chamber, where the temperature at the time of the impact was approximately $-50${$^{\circ}$}C. The strengths of the ice (both tensile and compressive, but to different degrees) increase with decreasing temperature. In our case,  the compressive and tensile strength of the targets was approximately 35 MPa and 3 MPa respectively, using data from \cite{propertiesice}. The porosity of our targets was measured to be $\textless$10\% and was determined by making a test sample of ice in an identical way to the targets in a  cubical box. The box was slightly under-filled with water so that a void remained at the top of the box after freezing. To measure this volume, a small amount of chilled ethanol (at -30$^{\circ}$C) was injected into the box. Since the mass and volume of the box are known, as well as the volume of injected ethanol and the temperature of pure water-ice (and hence density), the porosity can be calculated.

\subsection{Experimental set-up}
\label{expmethod}
The gun used to perform the experiments was the horizontal two-stage Light Gas Gun (LGG) of the University of Kent \citep{1999MeScT..10...41B}. It fires a shotgun cartridge in the first stage, which drives a piston to further compress a pressurised light gas in the pump tube. This gas is then suddenly released from its high pressure when a retaining disc of aluminium ruptures. This releases the gas into the second stage, where it accelerates the projectile. The projectile, which is placed in a sabot made of isoplast, is launched and travels down the gun range.
Two laser light screens are placed downrange and record the time of flight. The known separation of the two lasers, plus the time taken for the projectile to cross between the two laser screens, gives the speed (to within $\pm$0.2\%) of the projectile before it enters the chamber and impacts the target. It should be noted that since the publication of \cite{1999MeScT..10...41B}, the Impact Group has developed the ability to fire non-spherical projectiles such as, for example,  gem-stones (as used herein) and icy projectiles \citep{Price2013263}.

\begin{figure}
\includegraphics[width=1.\columnwidth]{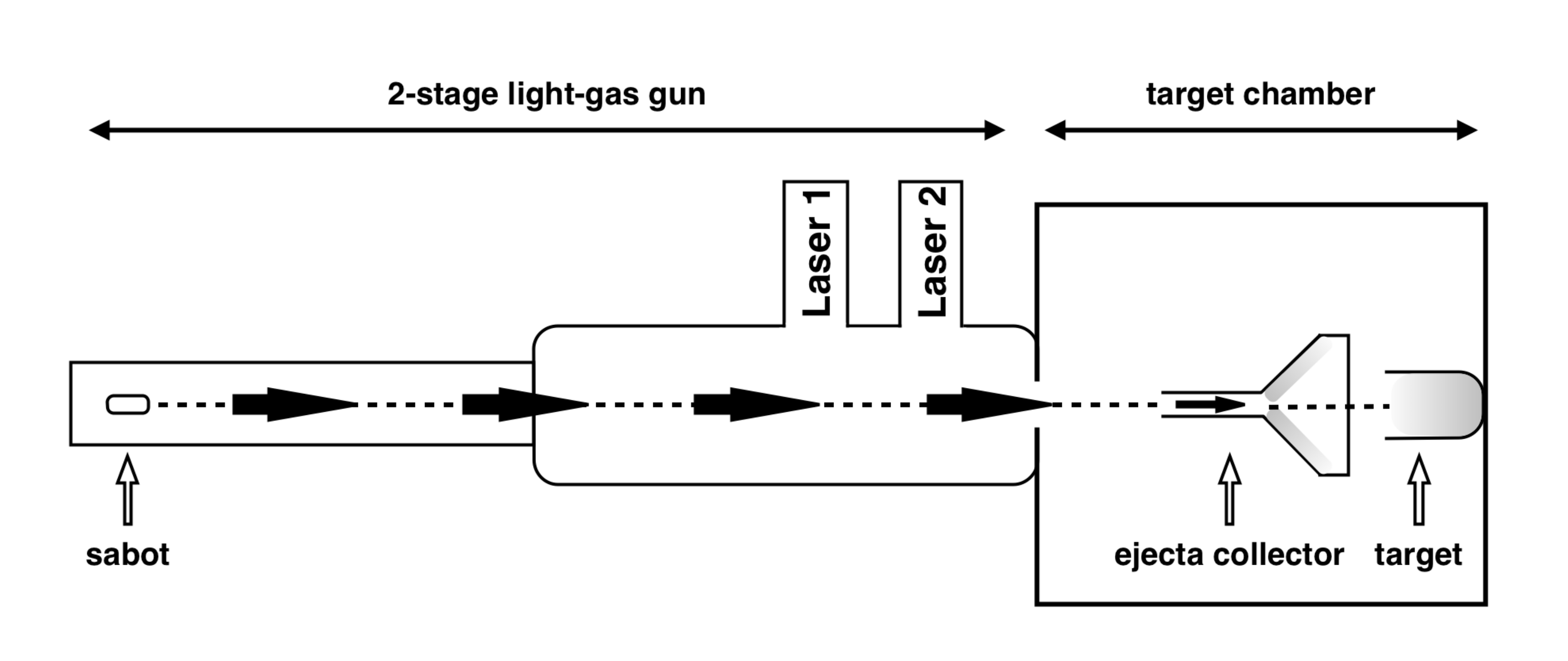}
\caption{The experimental set-up showing the projectile, which was placed inside a sabot, inside the two-stage LGG, and the configuration of the target chamber. The projectile impacts vertically the target at 0$^{\circ}$ in respect to its trajectory (dashed line).} The ejecta collection funnel was aligned with the flight path of the projectile and the centre of the target. It contained water-ice layers in order to collect the projectile's debris after the impact.
\label{setup}
\end{figure}

The pressure measured in the target chamber was no less than 50 mbar, due to the continuous sublimation of the ice target during the experiments. The impact angle was always  0$^{\circ}$. Here ‘zero degree’ is defined as impacting parallel to the impactor trajectory, and 90 degree to the target’s ambient plane. According to numerical simulations, which were applied to craters on the Moon's surface, the biggest proportion of the impactor's material remained in the crater for impacts occurring at 0$^{\circ}$ angles (see Fig.\ref{setup}). Decreasing amount of projectile material is expected to be embedded in the target with increasing impact speed, as has been demonstrated for the Moon's surface by \cite{2008LPI....39.2045B} and, more recently, by  \cite{2016Daly} and,  \cite{Dalynew} for asteroid surfaces.
In order to study this effect, we used impact speeds between 0.38 and 3.50 km/s. 

All projectiles were weighed and Raman spectra of the peridots were taken before each shot. These initial spectra were used as a comparison to examine the state of the largest fragment after the impact. Previous impact experiments have shown shifts in Raman spectra of the shocked target - and the magnitude of this shift has potential to be used as a shock 'barometer' \citep{2006GeCoA..70.6201K}. However, as the basalt has a glassy matrix it does not give well-defined, distinguishable, peaks in the Raman spectrum and no further spectra of the synthetic basalt projectiles was undertaken.

As one of the aims of this project was to measure the size distribution of the projectile's fragments after  impact at different speeds, we constructed a set-up to collect the ejecta (see Fig.\ref{setup}). A funnel with an internal water-ice layer was developed. 
The use of a water-ice coating led to a simple recovery technique of the ejecta fragments. However, secondary fragmentation is possible, and unavoidable using any sort of practical collection technique we can employ. In these experiments the secondary fragmentation is  minimal, as the speed of ejecta is only a small fraction of the impact speed \citep{2002holsappleast3,2012Burchell}. Additionally the ejecta fragment size is smaller than the projectile's size, and therefore less prone to fragmentation due to its smaller size. Finally, we do accept that these fragments have been shocked and weakened during the primary impact process.
As the projectile entered the target chamber it flew through the funnel, which completely covered the front of the target, and hit the centre of the target. Ejecta from the target was ejected and caught in the interior surface of the funnel. After each shot, the funnel was removed and the ice allowed to melt. In an identical way to the target, the melt ice was filtered and the majority of the projectile fragments were collected. Any ejecta that travelled backwards at small ejection angles (4.7$^{\circ}$$\pm$0.3) as measured from the projectile's trajectory, was able to escape the funnel, but were collected directly from the target chamber floor which had been covered before the shot with sheets of clean aluminium foil.

\subsection{Identification of fragments}
\label{fragments}

The first step after each shot was to search for the largest surviving fragment of the impactor. This was done by visually examining the crater in the target, the floor of the target chamber and the ejecta collector. Interestingly, for all the peridot shots except one (shot G260215 where the largest fragment was found in the target) the largest fragment was found on the target chamber's floor, implying that this largest fragment `bounced' backwards along its original flight path after impacting the target. For the spherical basalt projectiles all the largest fragments were found in the ejecta funnel, except from  shot G260515 were the largest fragment were recovered from the crater of the target. For each shot, the mass ( M$_{\mathrm{l,f}}$) of the largest recovered fragment is  measured with a balance with a precision of 10$^{-4}$ grams.

In order to identify the rest of the fragments which could not be visually inspected, we melted the ice and filtered the pure water from the target and the ejecta collection apparatus. These filters contained the projectile fragments mixed with contaminating material from the gun. These were fragments from the burst-disc, sabot, shotgun cartridge and any particulates picked up from the range during re-pressurisation of the target chamber. The majority of this material is C, Fe, Al and Si (see Fig.\ref{sex}a), but is a dust with a size (a few - 100s of microns) comparable to the projectile fragments we were interested in. That, currently unavoidable, contamination led us to develop a novel way to discriminate, count and measure the olivine fragments.

The effective area of each filter that contained the particles was a circle with a diameter of 37 mm. 
Images of the projectile's fragments were obtained by scanning the filters using a Back-Scattered Electron detector (BSE) on a SEM. Energy-dispersive X-ray spectroscopy (EDX) maps were taken of the same fields in order to distinguish projectile fragments from any contaminating material. We thus recorded information about the elemental composition of the sample. Considering that; (i) the peridot projectiles have a very strong Mg signal and, (ii) there is very little Mg contamination from gun debris, we used the EDX maps of Mg to discriminate the projectile fragments (see Fig.\ref{sex}b from contaminating gun debris).

\subsection{Estimation of projectile fragmentation}

The energy density has long been used to assess disruption of  projectiles \citep{1979davis,1990schultzgsa}. In this work, 
following  \cite{2014M&PS...49...69N}, the energy density at the time of the impact is Q (J/kg), and its form for the impactor is given by:

\begin{equation}
Q_{\mathrm{im}}=\frac{1}{2} v^{2}
\end{equation}
where $\upsilon$ (m/s) is the impact speed.
Traditionally, it is assumed that catastrophic disruption occurs when M$_{\mathrm{l,f}}$/M$_{\mathrm{im}} \leq$ 0.5, with the energy threshold of Q*.
Plots of M$_{\mathrm{l,f}}$/M$_{\mathrm{im}}$ vs. Q$_{\mathrm{im}}$ are used to give an estimate of the projectile fragmentation as a function of the impact velocity (energy).  

\subsection{Determination of the SFDs of the ejecta projectile fragments}
\label{sextractor}

Another quantity that gives crucial information about the fragmentation of the projectile is the size frequency distribution (SFD) of the fragments: for instance, steep cumulative SFDs are indicative of projectiles being pulverised by the impact, whereas shallow cumulative SFDs indicate that large fragments coexist with small ones. Moreover, the size at which the differential SFD has peaks (or a peak) indicate the typical dimension of the fragments. These peaks are also called fragmentation modes.

\begin{figure*}
  \includegraphics[width=1\textwidth]{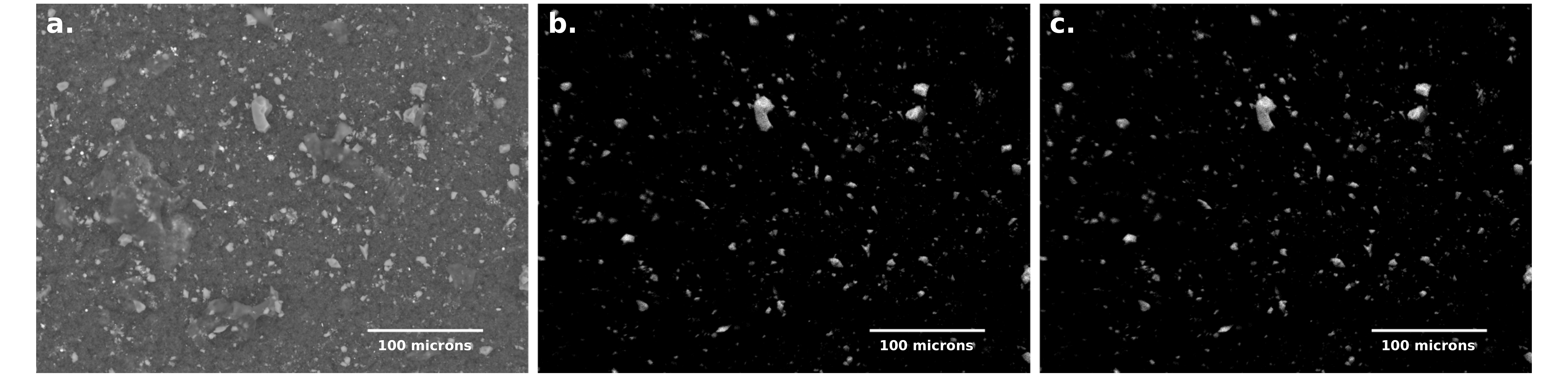}
  \caption{SEM image showing that the fragments of the projectile are mixed with other material  from the gun (a). As the projectile is Mg-rich it gives a strong signal in Mg X-ray maps (b). Considering also that there are no other sources of Mg contamination, these maps are used as the main dataset for the analysis. SEXTRACTOR identifies each fragment and reproduces another image containing only the pixels which contain information according to the given threshold (c).}
\label{sex}  
\end{figure*}

SFD calculations were made by measuring the sizes of the impactor fragments stopped by, and accumulated on, the ejecta collector. Once the ejecta collector ice was melted and the fragments were collected on polytetrafluoroethylene (PTFE) filters (pore size 0.1 $\umu$m), we acquired two maps per filter, consisting of 50 SEM and 50 EDX frames, the latter required 30 minutes acquisition time per frame. Each frame contained hundreds of fragments (see Fig.\ref{sex}b) and was taken with a magnification of $\times$300,  giving a pixel scale of 0.4 $\umu$m/pixel, which enabled us to detect even very small fragments. Manually counting the fragments and measuring their dimensions is extremely time consuming and prone to observer bias. To tackle this, we applied an astronomical photometry technique to each image using the SOURCE EXTRACTOR ('SEXTRACTOR')  open source software for astronomical photometry \citep{1996A&AS..117..393B}. 
SEXTRACTOR is a program specifically written to automatically identify and measure extended light sources, such as galaxies, from astronomical images. To prepare the SEM-EDX images to be suitable for use by SEXTRACTOR, we converted the raw data to 16-bit Flexible Image Transport System (FITS) files, making sure that there was no loss of information through the transformation.  Unlike most galaxies, which are well defined elliptical sources, mineral fragments are irregular in shape. Therefore, to measure the total X-ray emission from a fragment, we used the ISO photometry setting within SEXTRACTOR, which is able to identify the shape irregularity of each fragment. 

As the background noise from the EDX images was very close to zero counts, we were able to set a very low detection threshold in units of the background's standard deviation. By selecting pixels with counts at least three times above the mean background noise, we were able to identify the vast majority of the fragments per field. An additional threshold for the minimum detected area was defined in order to increase the detection reliability. SEXTRACTOR measures the semi-major and semi-minor axes allowing each object to be described as an ellipse.
According to the threshold, which constrains the size of the minimum area identified as a fragment, SEXTRACTOR reproduces another image containing only the identified fragments, as shown in Fig.\ref{sex}, not measuring anything smaller. By examining the new images we verified that there was no false detections due to background noise.

If the field is very crowded with fragments, there is the possibility of blending the X-ray emission of several fragments. SEXTRACTOR comes with a sophisticated de-blending algorithm which flags the initially blended fragments.  
SEXTRACTOR  has the ability to discern shapes even in highly dense fields, giving good statistics by automatically counting thousands of fragments. SEXTRACTOR also has an edge detection algorithm and ignores fragments that lay on the edge of an image.  
However in order to avoid false detections due to noise, we set SEXTRACTOR to identify minimum fragment areas of 0.64 {$\umu$m}$^{2}$.

\subsection{Estimation of projectile material in the target}

We used the same approach described in Section \ref{sextractor}, to analyse the filters of the target melt water. These filters collected projectile material once the target ice was melted and filtered away. 
While mapping the target filters, in contrast with the mapping of the ejecta filters, we used low magnification ($\times$50) in the SEM. This is because we noticed some spatial variability in the number of fragments on the filter and we choose to map the entire surface of the filter to detect all possible impactor fragments. The chosen resolution enabled us to scan a whole filter in approximately 24 hours with pixel scales between 4.4 -- 4.9 $\umu$m/pixel and thus create a mosaic of the whole filter area. This means that if the fragments that remained in the target follow a similar size-frequency distribution with the ejecta fragments, then we should expect to have an amount of fragments smaller than a pixel. 
The significant factor to consider in choosing a detection threshold in SEXTRACTOR is the background noise of the images, which is due to the Bremsstrahlung radiation as the electron beam decelerates within the sample. The level of this background noise is different for each different element. Ideally, if there was no Bremsstrahlung background we could use an extremely low threshold for the minimum detected fragment area as every pixel with value greater than 0 corresponds to a real Mg signal. That way we could measure fragments with sizes as small as the pixel scale. However this is not possible and, in order to overcome the problem, we performed the analysis of the maps using SEXTRACTOR choosing several different thresholds for the minimum detected area.

After having extracted the 2D area of each fragment, as projected on the X-ray detector, an extra step was performed in order to estimate a z-length that corresponds to the fragment's height. As there was not a preferable position of the fragments we were therefore able to adopt simple estimations of the z-axis which was assumed to follow the same distribution of x and y axes. Several studies so far, when an estimation of a volume was demanded, use simple formulae to estimate the z-axis dimension; such as a simple average of the x and y dimensions. 

Considering that the produced fragments are cubic-shaped, we estimate in Table \ref{table2} the total mass of the projectile by Eq.\ref{mass}:
\begin{equation}
M_{\mathrm{p}}= \sum_{i=0}^N x_i \times y_i \times \frac{(x_i+y_i)}{2} \times \rho 
\label{mass}  
\end{equation}
where x$_{i}$ and y$_{i}$ are the big and small axis of each fragment respectively, and $\rho$ = 3.217 gr cm$^{-3}$, the density of the peridot.

\section{Results}
\label{results}

\subsection{State of the largest projectile surviving fragments}
\label{state}

\begin{figure}
\includegraphics[width=1\columnwidth]{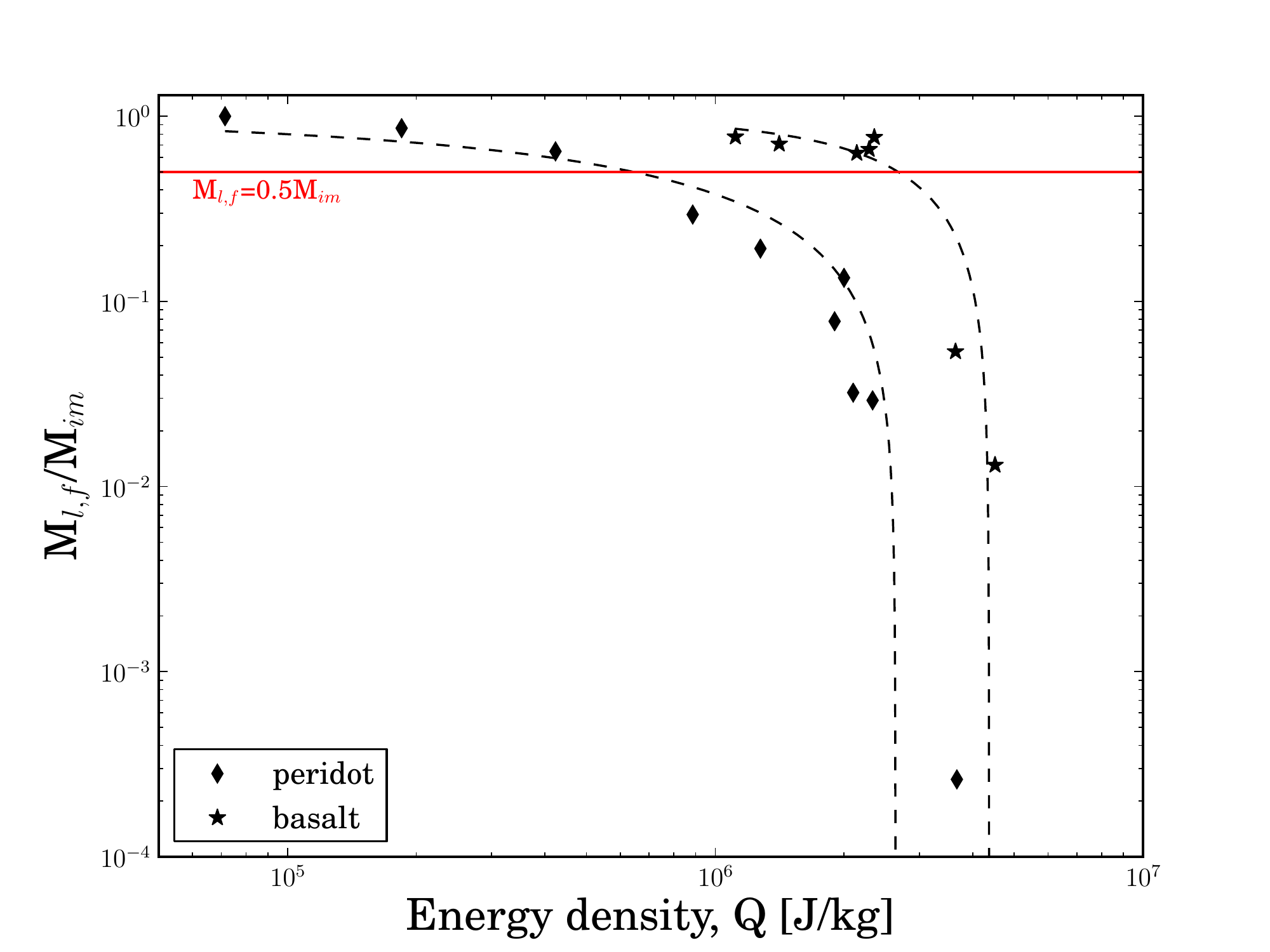}
\caption{Mass ratio of the largest surviving fragment of the impactor versus the energy density, Q$_{\mathrm{im}}$, for speed ranges 0.38 -- 2.71 km/s and 1.49 -- 3.03 km/s for olivine and basalt respectively.  The dashed lines correspond to the best-fitting curves using Eq.\ref{eq:Q}.}
\label{qplot}
\end{figure}

\begin{figure}
\includegraphics[width=1\columnwidth]{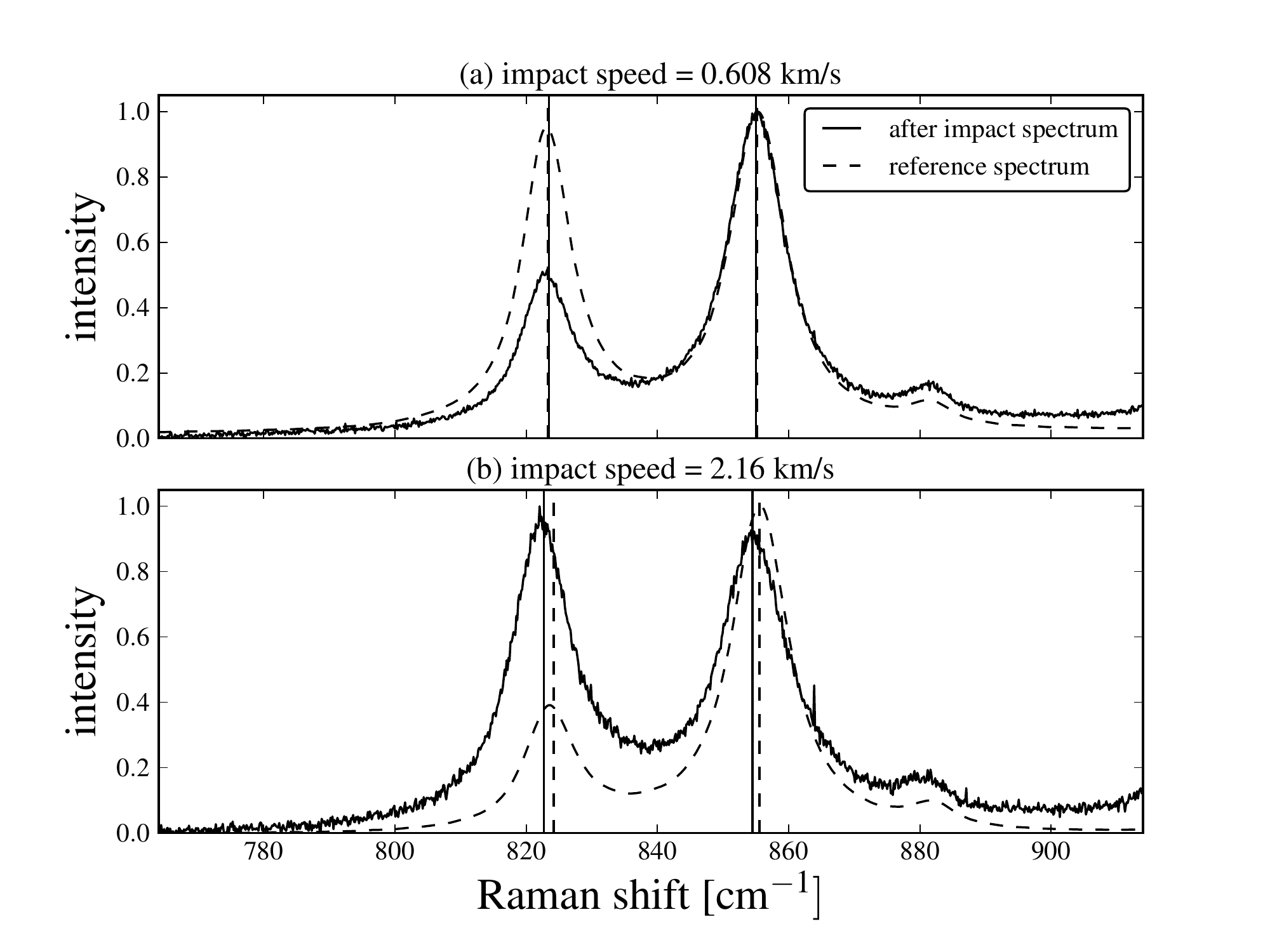}
\caption{At 0.92 km/s none of the shifts exceed the precision of the instrument,  whilst for the 2.16 km/s shot a shift in  P$_1$ and P$_2$ olivine lines was observed to be 1.49 cm$^{-1}$ and 1.08 cm$^{-1}$ respectively.}
\label{both}
\end{figure}

In Fig.\ref{qplot} we present the mass of the largest fragment we retrieved as a fraction of the initial impactor's mass, in relation to  the energy density Q$_{\mathrm{im}}$. 
In order to calculate the values of the energy density at the catastrophic disruption threshold, Q*$_{\mathrm{im}}$, we fit the parameters of Eq.\ref{eq:Q} to the data:
\begin{equation}
\frac{M_{\mathrm{l,f}}}{M_{\mathrm{im}}} = 1-A Q_{\mathrm{im}}^{c}
\label{eq:Q}
\end{equation}
 We found that c$_{\mathrm{p}}$= 0.49,  A$_{\mathrm{p}}$= 6.80$\times$10$^{-4}$ for peridot and c$_{\mathrm{b}}$= 1.50, A$_{\mathrm{b}}$=1.42$\times$10$^{-10}$ for basalt fragments.
The derived values of the catastrophic disruption threshold, Q*$_{\mathrm{im}}$, were estimated at 7.07$\times$10$^{5}$ J/Kg and 2.31$\times$10$^{6}$ J/Kg for peridot and basalt respectively.\\
Raman spectra of the recovered fragments, using a near-IR laser at 785 nm, were obtained to ascertain whether the impact shock caused a shift in the main olivine lines, referred to as P$_1$ and P$_2$.  The P$_1$ and P$_2$ lines are at 822.64 -- 824.20  cm$^{-1}$ and 854.15 -- 855.63 cm$^{-1}$ respectively at the reference spectra of the projectiles which were measured before each shot  \citep{2014EPSC....9..295H}. By comparing the spectra we collected before, and after, each shot we noticed a small shift of the two prominent olivine lines which slightly increased with increasing collisional speed, as shown in Fig.\ref{both}. The greatest shift measured was 1.49 and 1.08 cm$^{-1}$ for the  P$_1$ and P$_2$ respectively, which we interpret as not significant since the accuracy of the measurement is approximately 1 cm$^{-1}$. 

\begin{figure}
  \includegraphics[width=1\columnwidth]{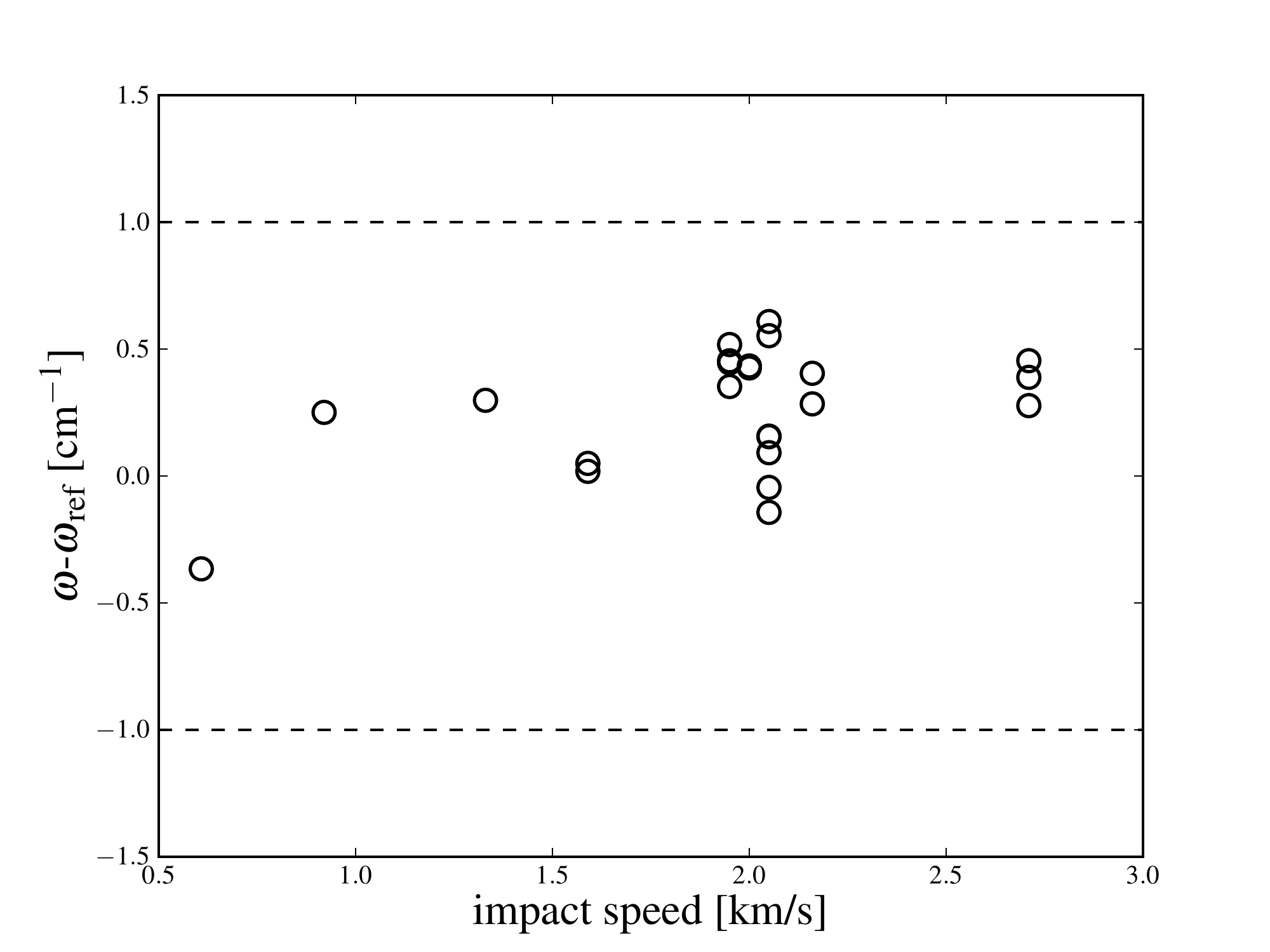}
\caption{The change is separation, $\omega$, of the P$_1$ and P$_2$ olivine lines was calculated for all the big surviving fragments in the range of impact speeds 0.608 -- 2.71 km/s.}
\label{diff}
\end{figure}

Another interesting application of Raman spectra would be the identification of any change in the separation ($\omega$) of the two characteristic peaks of forsterite which, together with elementary quantification of Mg and Fe, could show possible shock induced change to the crystallisation and/or the elemental composition of the olivine \citep{2006GeCoA..70.6201K, Foster20131}. These two prominent peaks are the result of the fundamental vibration of the chemical bonds (here of Si-O bonds). Peak positions and shifts are generally used to calculate the ratio of Mg/(Mg+Fe) in olivine. The positions of the P$_1$ and P$_2$ are strongly related to Fe and Mg compositions of the olivine. For example according to \cite{2006GeCoA..70.6201K} the separation of the P$_1$ can be up to 10 cm$^{-1}$ from fayalite  to forsterite while the separation of the P$_2$ can be up to 20 cm$^{-1}$.
Up to our maximum collision speed, no change in $\omega$ was detected above the spectral resolution of the spectrometer ($\sim$1 cm$^{-1}$) (Fig.\ref{diff}).

\subsection{Impact strength}
\label{strength}

In order to investigate the peak pressures and temperatures experienced by the projectile during impact, a complementary program of hydrocode modelling was undertaken. 

Simulations were performed with the AUTODYN  hydro-code \citep{Hayhurst1997337}. A simple Lagrangian, 2-D half-space model was set up, using 20 cells across the projectile's radius. The total number of cells in the model was approximately 500,000. Material models for ice were taken from \cite{2013AdSpR..52..705F} using a 5-Phase Equation-of-state (EoS) from \cite{2011Icar..214...67S}. Strength and EoS data were taken from \cite{PG2012120} and  \cite{mat_properties} respectively. Gauges (or tracers) were placed along the axis of the projectile so that pressure and temperature could be determined during the impact. In Table \ref{table1} we present the peak pressures, P$_{\mathrm{max}}$, the temperatures at the time of the peak pressures, T$_{\mathrm{P}}$,  was experienced and the maximum temperature, T$_{\mathrm{max}}$, 1 $\umu$m below the front surface of the projectile.  
Note, that  for the lowest speed shot (0.38 km/s) the peak pressure as modelled does not exceed the yield strength of olivine (1.5 GPa). This agrees with the observed state of the recovered projectile, that retained 100\% of its initial mass and showed no signs of damage. 

\begin{table*}
\centering
\begin{tabular}{lccccc|lccccc}
\hline
Peridot & speed  &  P$_{\mathrm{max}}$ &  T$_{\mathrm{max}}$  &  T$_{\mathrm{P}}$ &  M$_{l,f}$/M$_{\mathrm{im}}$ & Basalt & speed &  P$_{\mathrm{max}}$ &  T$_{\mathrm{max}}$ &  T$_{\mathrm{P}}$ & M$_{l,f}$/M$_{\mathrm{im}}$\\
\hline
Shot No &  [km/s] &  [GPa] & [K] & [K] &  [\%] & Shot No &  [km/s] &  [GPa] & [K] & [K] &  [\%] \\
\hline
S141114  & 0.38  &  0.54 & 301 & 293 &  100 &G010415&1.49&0.80& 360 &307& 77.42 \\
S180315  & 0.60  &  1.21  & 298 & 293 &  86.25 &G260515&1.68&1.02& 401 &302& 70.93\\
S211114  & 0.92  &  1.64  & 297 & 294  &  65.50 &G240415&2.07&1.29& 433 &303& 63.20   \\
G060315  & 1.33  &  2.84  & 302 & 295   &  29.46 &G050615&2.14&1.32& 436 &303& 66.32  \\
G260215  & 1.60  &  3.75  & 312 & 296  & 19.30 &G260515&2.17&1.33& 440 &303& 77\\
G230115  & 1.95  &  4.83  & 330 & 297  & 7.80 &G070515&2.70&2.97& 463 &308& 5.36\\
G250315  & 2.00  &  4.94  & 331 & 297  & 13.42  &G270415&3.03&4.58& 522 &317& 1.30\\
G261114  & 2.05  &  5.06 & 342 & 297  & 3.21 &&&&&&  \\
G130315 & 2.16 &  5.59  & 335 & 298  & 2.92  &&&&&& \\
G180215 & 2.71 &  7.13 & 397 & 299  & 0.02  &&&&&& \\
G031214 & 2.97 &  8.04  & 407 & 305  &-&&&&&& \\
G121214 & 3.50 &  10.2  & 513 & 353  &-&&&&&&  \\
\hline
\end{tabular}
\caption{ Peak pressure, P$_{\mathrm{max}}$, peak temperature, T$_{\mathrm{max}}$, and temperature at peak pressure, T$_{\mathrm{P}}$ , are shown for the range of shots at the time of the impact. M$_{\mathrm{l,f}}$/M$_{\mathrm{im}}$ represents the proportion of the largest fragment of the impactor of its initial mass. For the shots G031214 and G121214 we were not able to identify the largest fragment. For the shots G260215 and G260515 the largest  fragments were recovered from the bottom of the craters.}
\label{table1}
\end{table*}

\subsection{SFDs of the ejecta projectile fragments}
\label{sizes}

\begin{figure}
  \includegraphics[width=1\columnwidth]{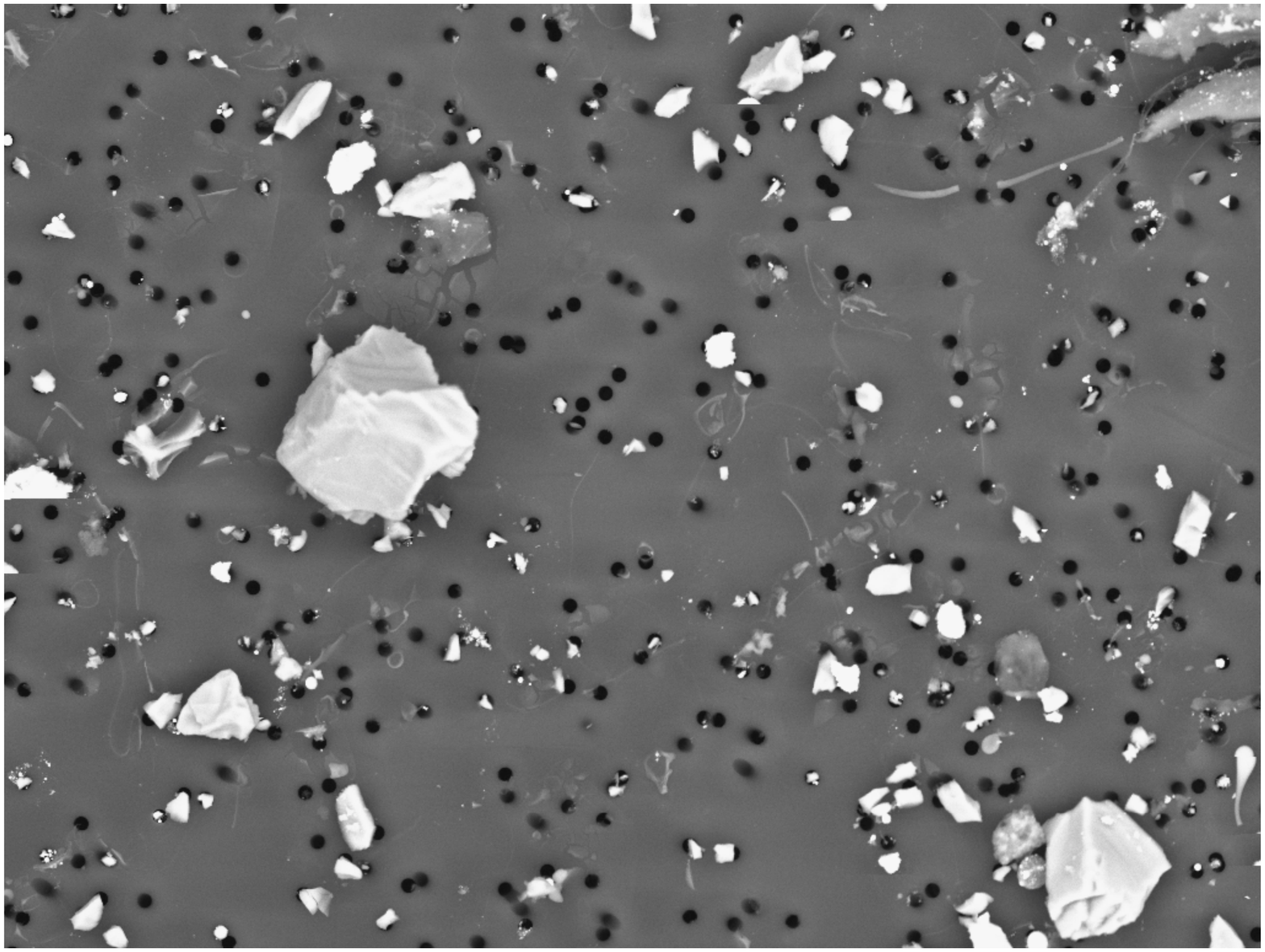}
  \caption{SEM image showing that a significant amount of olivine fragments smaller than 0.1 $\umu$m, which is the pore size (black circles), remained on the filter.}
\label{filter}  
\end{figure}

Following the procedures described in Sections \ref{fragments} and \ref{sextractor}, we measured the fragment SFDs for all our shots. 
 A noticeable number of fragments smaller than 0.1 $\umu$m remained on the filter lying between the holes (see Fig.\ref{filter}), but as the resolution of the SEM-EDX images was 0.4 $\umu$m per pixel, we were not able to measure fragments  smaller than the resolution using our automated image analysis routines. Therefore, $\sim$0.4 $\umu$m is, effectively, the limiting spatial resolution of our SEM. 

\begin{figure}
  \includegraphics[width=1\columnwidth]{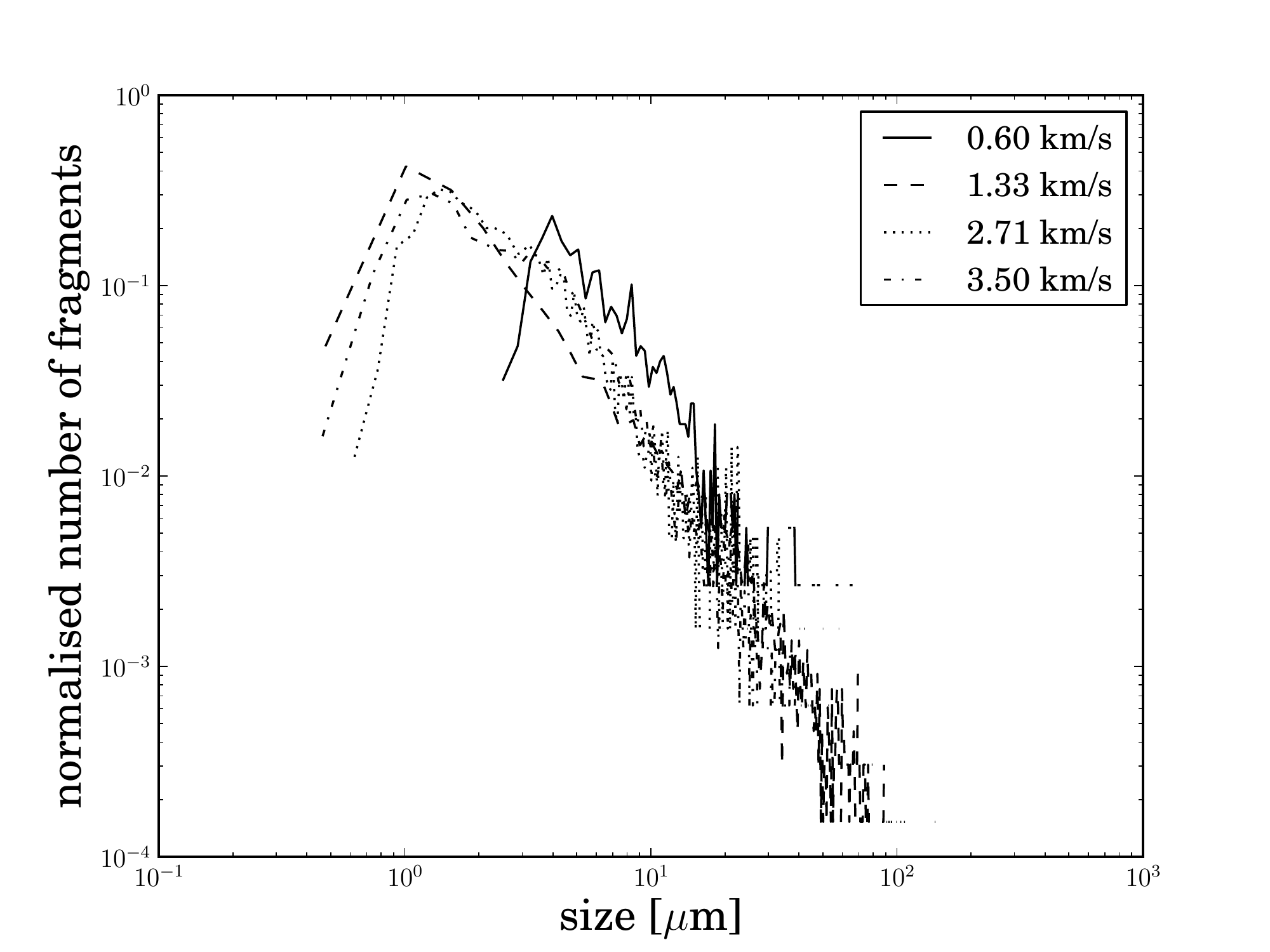}
  \caption{Size frequency distributions of indicative shots, showing no significant change in modes. }
\label{dist}  
\end{figure}

\begin{figure}
  \includegraphics[width=1\columnwidth]{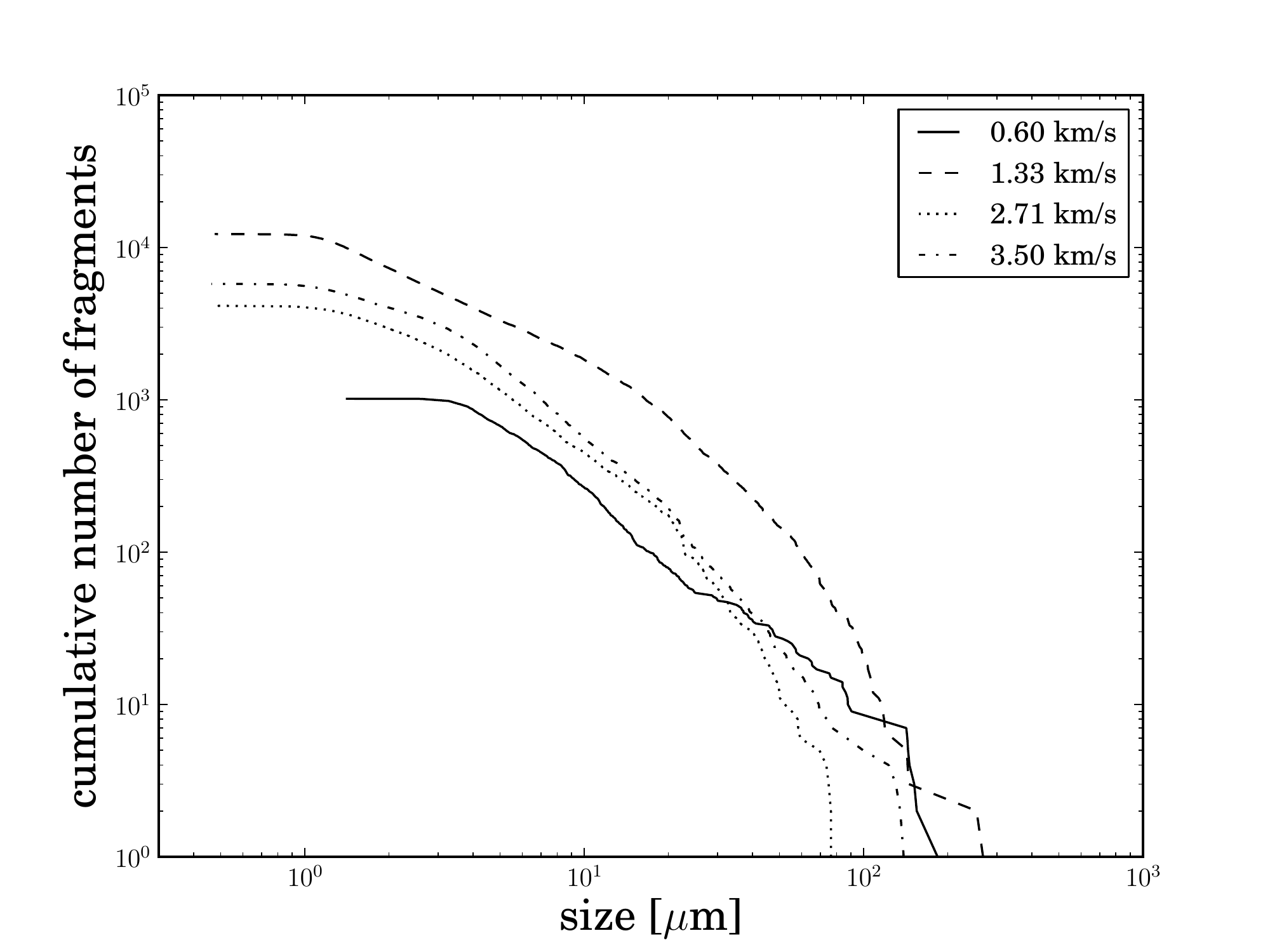}
  \caption{Cumulative size frequency distribution of the same shots shown in Fig.~\ref{dist}, demonstrating that the fragmentation of peridot does not change by increasing collision speed.}
\label{dist_cum}  
\end{figure}

The SDFs of the size of the fragments appear to have a power-law tail, as shown in Fig.\ref{dist}. There is a shift of approximately 3 $\umu$m of the principal mode of the distribution from 0.608 to 1.33 km/s shots but beyond this speed the mode remains constant at around 1.5 $\umu$m. Considering the size of the filters (0.1 $\umu$m) and the detection threshold of the EDX maps (0.4 $\umu$m), the turnover of the curves around 2 $\umu$m is real, and not an artefact of the detection process. 
It would be expected that as the impact speed increases the number of smaller fragments would increase.  However, as can be seen from  Fig.\ref{dist} and Fig.\ref{dist_cum}, although there are differences of even an order of magnitude  in the number of fragments, there is no clear trend in the fragmentation behaviour with increasing impact speed. Similarly the slopes of the cumulative distributions in Fig.\ref{dist_cum} also show no clear trend with increasing speed, which seems to be counter-intuitive. 
We found that the slopes of all size frequency distributions lie in a range between -1.04 and -1.68.
Here we have to point out that due to a possible secondary fragmentation that occurred on the ejecta collecting system the observed slopes of the SFDs would be steeper. However, as described in Section \ref{expmethod}, we expect this phenomenon to be limited.

\subsection{Implantation of material in the target}
\label{contamination}

In Table \ref{table2} we present the overall masses of the fragments which were found in the target filters as a fraction of the initial projectile mass. For the identification of the fragments we used two different photometric thresholds for the minimum detected area (4 and 6 pixel area respectively). Note that for the shot G260215 the largest fragment was found in the crater with a mass of 31\% of the initial mass, increasing the total amount implanted in the target from 53 -- 55\%.

\begin{table}
\centering
\begin{tabular}{lccc}
Shot & speed [km/s]  &  area 4 [\%]  & area 6 [\%]\\
\hline
S180315  & 0.60  & 0.2 & 0.18\\
S211114  & 0.92  &  0.43 & 0.37\\
G060315  & 1.33  &  1.4 & 1.17\\
G260215  & 1.60  &  24 & 22.4\\
G230115  & 1.95  &  8.29 & 0.58\\
G250315 & 2.00 & 1.71 & 0.55\\
G220515 & 2.04 & 1.15 & 0.48\\
G130315 & 2.16 &  3.5 & 2.6\\
G180215 & 2.71 &  0.17 & 0.12\\
\hline
\end{tabular}
\caption{The mass fraction of the olivine projectile that was found embedded in the targets using two different detection thresholds, of 4 and 6 pixels respectively, for the minimum detected area. }
\label{table2}
\end{table}

\section{Discussion}
\label{discussion}

At impact speeds up to 3 km/s, which occur at the lower part of the velocity distribution in the Main Belt \citep{1994Icar..107..255B,2011obrien}, we observe no detectable melting of the projectile, as determined by visual observation and Raman spectroscopy (melting of olivine results in a degradation of the Raman spectra due to loss of olivine crystal structure). This observation is backed up by hydrocode modelling which demonstrates that the  temperatures at maximum pressures (Table \ref{table1}) experienced by the olivine and basaltic impactors do not reach their melting point, which is 2100 K and 2500 K at pressures of 10 GPa and 4.6 GPa respectively. This is an important observation when we consider the mineralogical signature of implanted impactors on asteroids i.e. the projectile's mineralogy (including crystallinity) will be preserved. 
By examining the Raman spectra of the survived fragments, and calculating the difference  $\omega$ - $\omega$$_{\mathrm{ref}}$, we found that there is no alteration in the Fe abundance of the fragments as all the calculated differences lie inside the resolution limit of the instrument. However it would be extremely important to identify the impact speed at which the olivine starts to melt and the introduced shock is enough to change the molecular geometry in the crystal.

The size distribution of the projectile fragments has a definite turn-over at a point well above the detection limit of our method. The positions of the modes and slopes of the size distributions are velocity invariant, although there is a difference between the modes at 0.608 and 1.331 km/s. This is counter to the observations made for ductile (i.e. metal) projectiles by \cite{Hernandez20061981},  \cite{2013kenkmann} and McDermott et al. (in preparation). This suggests that the fracturing mechanism between lithological projectiles (non-ductile) and metallic (ductile) projectiles is different.

We determine different Q* values, with an order of magnitude difference, for forsterite olivine and synthetic basalt onto icy surfaces at speeds relevant to impacts in the asteroid belt. Comparing the results of both projectiles it is obvious that the same portion of mass of the basaltic projectile, survives at higher collisional speed than the peridot. Additionally, the data demonstrate that significant fractions of projectile material survives and escapes as ejecta. The main result  is that at collision speeds close to 3 km/s there is material implanted in the target even if its mass is only a small proportion of the initial mass of the projectile. The data points to that the portion of the mass implanted in the target is related to the type of materials which collide, and the porosity of the target. As the porosity of the same material increases, the compressive strength decreases and this, in turn, affects the result of the impact. Higher porosity leads to the formation of narrower and deeper craters because the shock-wave cannot propagate as easily as in the non-porous materials, and the energy is concentrated in a limited cross-section area. Moreover, the ejecta velocities are reduced as the porosity of the target increases, even up to two orders of magnitude, and therefore there are indications that larger amount of material will be eventually implanted in the target from the re-accumulation of the ejecta (containing projectile debris).  There are already several examples from laboratory experiments on highly porous targets, trying, among others, to simulate collisions on 253 Mathilde with porosity $\sim$50\% \citep{1999housen}, that show very limited or even no ejecta material is found around the crater. The implications of this, along with our current results and ongoing experiments, can contribute to the explanation of the formation of multi-lithology small bodies, considering also ejecta velocities smaller than the escape velocities of these bodies. A new set of ongoing experiments may prove this hypothesis. In these experiments peridot and basaltic projectiles, of the same sizes and strengths as in the described experiments, are being fired at water-ice targets with porosity between 35 -- 40\% which is similar to the average bulk porosity of C-type asteroids. 

Finally, It should be noted that the results presented here are for near normal impacts only, and that the observations may differ as a function of impact angle. \cite{2005Schultz} and  \cite{2007Schultz}, who were investigating the Deep Impact impact, demonstrated that the cratering mechanism differs between normal and oblique impacts. In addition, recent work from \cite{2016Daly} investigated the mass implanted in a target at different impact angles and showed that the embedded projectile material is reduced with increasing impact angle.

\section{Conclusions}

Our experiments have demonstrated a difference in the fragmentation of the forsterite olivine and synthetic basalt projectiles, that were fired onto low porosity water-ice targets, giving catastrophic disruption energy densities of Q*$_{\mathrm{p}}$ = 7.07$\times$10$^{5}$ J/Kg and Q*$_{\mathrm{b}}$ = 2.31$\times$10$^{6}$ J/Kg respectively. We note that there is no change in modes and slopes of the SFD of the olivine beyond the impact speed of 1.331 km/s. In addition we did not record any melt or vaporisation of the projectile for the range of impact speeds 0.38 -- 3.50 km/s. Therefore we suggest that, for such velocities that represent the lower end of the distribution of impact velocities in the main asteroid belt (about 5 km/s),  there should be significant survival of the impactors.

In this work we also present a novel way to measure thousands of fragments autonomously and accurately, in order to study the fragmentation properties of the projectile during a hypervelocity impact with unprecedented statistical significance. Applying astronomical photometry techniques enabled us to measure fragments down to sizes of a few microns, and adequately define the 2D area (and thus inferred volume) of each fragment. This analysis method is essential to estimate SFD and masses of very small fragments, as the LGG cannot fire bigger projectiles, which will produce larger fragments suitable for weighing. 

\section*{Acknowledgments}
CA would like to thank the University of Kent for her 50$^{th}$ Anniversary PhD scholarship and Timothy Kinnear and Victor Ali-Lagoa for fruitful discussions. MCP and MJC thank the STFC, UK for funding this work. MD acknowledges support from the French Agence National de la Recherche (ANR) SHOCKS.

\bibliographystyle{mn2e}
\bibliography{references}

\label{lastpage}

\end{document}